 \documentstyle[emulateapj,flushrt,epsf,rotate]{article}

\def\bt{\begin{tabbing}}
\def\et{\end{tabbing}}
\def\beq#1{\begin{equation}\label{#1}}
\def\eeq{\end{equation}}

\def\fl1{FL1}
\def\fl2{FL2}
\def\ie{$i.e.$}
\def\eg{$e.g.$}
\def\etal{et al.~}

\def\yi{$y_i$}
\def\bfy{{\bf y}}
\def\bfyT{\tilde{\bf y}}

\def\xi{$x_i$}
\def\bfx{{\bf x}}
\def\bfxt{{\bf x}^t}
\def\bfxT{\tilde{\bf x}}
\def\bfxTt{{\tilde{\bf x}}^t}

\def\bfri{{\bf \hat{r}}_i}

\def\bfA{{\bf A}}
\def\bfAt{{\bf A}^t}
\def\bfAT{\tilde{\bf A}}

\def\bfn{{\bf n}}
\def\bfnt{{\bf n}^t}
\def\bfnT{\tilde{\bf n}}
\def\bfnTt{\tilde{\bf n}^t}

\def\bfM{{\bf M}}
\def\bfN{{\bf N}}

\def\bfNT{\tilde{\bf N}}

\def\bfW{{\bf W}}
\def\bfWt{{{\bf W}}^t}
\def\bfI{{\bf I}}
\def\bfD{{\bf D}}
\def\bfDt{{{\bf D}}^t}
\def\bfS{{\bf S}}

\def\eta{\varepsilon}
\def\etat{{\varepsilon}^t}
\def\Sig{{\bf\Sigma}}

\def\bfrh{\hat{\bf r}}

\def\microK{\mu{\rm K}}
\def\sn{$S/N$}
\def\bfb{{\bf b}}
\def\bfbt{{\bf b}^t}
\def\bfB{{\bf B}}
\def\bfBt{{\bf B}^t}
\def\bfz{{\bf z}}

\def\expec#1{\langle#1\rangle}
\def\tr{\hbox{tr}\>}
\def\l{\ell}
\def\etal{{\frenchspacing\it et al.}}
\def\ie{{\frenchspacing\it i.e.}}
\def\eg{{\frenchspacing\it e.g.}}
\def\etc{{\frenchspacing\it etc.}}

\def\spose#1{\hbox to 0pt{#1\hss}}
\def\simlt{\mathrel{\spose{\lower 3pt\hbox{$\mathchar"218$}}
     \raise 2.0pt\hbox{$\mathchar"13C$}}}
\def\simgt{\mathrel{\spose{\lower 3pt\hbox{$\mathchar"218$}}
     \raise 2.0pt\hbox{$\mathchar"13E$}}}
\def\simpropto{\mathrel{\spose{\lower 3pt\hbox{$\mathchar"218$}}
     \raise 2.0pt\hbox{$\propto$}}}

\def\pp{\noindent\parshape 2 0truecm 13.6truecm 1truecm 12.6truecm}
\def\rn{\pp}

\def\bfig{\begin{figure}[h] \centerline{\hbox{}}\vfill}
\def\efig{\end{figure}\vfill\newpage}

 \journalid{337}{15 January 1989}
 \articleid{11}{14}

\slugcomment{Submitted to ApJL August 5, 1998}
\lefthead{de Oliveira-Costa {\etal}}
\righthead{MAPPING THE CMB}
\begin{document}

\title{Mapping the CMB III: combined analysis of QMAP flights}

\author{Ang\'elica de Oliveira-Costa$^{1,2}$, 
        Mark J. Devlin$^{3,1}$, 
        Tom Herbig$^1$, 
        Amber D. Miller$^1$,}
\author{C. Barth Netterfield$^{4,1}$, 
        Lyman A. Page$^1$ \& 
        Max Tegmark$^{2,5}$}
\affil{$^1$Princeton University, Department of Physics, Jadwin Hall, 
      Princeton, NJ 08544; angelica@ias.edu}
\affil{$^2$Institute for Advanced Study, Olden Lane, Princeton, 
      NJ 08540}
\affil{$^3$University of Pennsylvania, Department of Physics and Astronomy,
      David Rittenhouse Laboratory, Philadelphia, PA 19104}
\affil{$^4$California Institute of Technology, MS 59-33, Pasadena, CA 91125}
\affil{$^5$Hubble Fellow}

\begin{abstract}
We present results from the QMAP balloon experiment, which maps
the Cosmic Microwave Background (CMB)
and probes its angular power spectrum on degree scales.  
In two separate flights, data were taken in six channels at two frequency bands 
between 26 to 46 GHz. 
We describe our method for mapmaking (removal of $1/f$-noise and 
scan-synchronous offsets) and power spectrum estimation, as well as 
the results of a joint analysis of the data from both flights.
This produces a 527 square degree map of the CMB around the 
North Celestial Pole,
allowing a wide variety of systematic cross-checks. 
The frequency dependence of the fluctuations is consistent with CMB 
and inconsistent with Galactic foreground emission. 
The anisotropy is measured in three multipole bands from $\ell\sim 40$ to 
$\ell\sim 200$, and the angular power spectrum 
shows a distinct rise which is consistent with the Saskatoon results. 
% Possible systematic errors are discussed.
\end{abstract}


%%%%%%%%%%%%%%%%%%% TEXT: %%%%%%%%%%%%%%%%%%%%%%%%%%%%%%%

\section{INTRODUCTION}

% The main goal of the new generation of Cosmic Microwave Background (CMB) 
% experiments is to produce maps that can resolve features as well as 
% reveal the shape of the angular power spectrum.

The QMAP balloon experiment was designed to map the Cosmic Microwave Background 
(CMB) and measure the angular power spectrum on degree scales. 
QMAP operates in the Ka ($\sim$ 30 GHz) and Q-band ($\sim 40$ GHz)
with six detectors in two polarizations
(Ka1 and Ka2; Q1 and Q2; Q3 and Q4),
with angular resolution between $0\fdg6$ and $0\fdg9$.
% The single Ka-Band and the two Q-Band feed horns were each split into two
% linear polarizations (Ka1 and Ka2; Q1 and Q2; Q3 and Q4).
Data were taken during two flights in 1996, the first (hereafter FL1) in June in 
Palestine, Texas, and the second (hereafter FL2) in November in Ft. Sumner, New Mexico. 
QMAP scanned a 527 square degree region near the North Celestial Pole in a complicated 
criss-cross pattern which allows a number of internal checks of the integrity of the 
measurements and results in an interconnectedness between pixels that enables efficient 
$1/f$-noise removal. 
% A frequency span of 20 GHz provides discrimination against foreground contaminants. 

A detailed description of the 
design and performance of the 
QMAP instrument and results from the first flight 
are presented in Devlin {\etal} (1998, hereafter D98). 
QMAP calibrations and results from the second flight are presented in 
Herbig {\etal} (1998, hereafter H98).
In this {\it Letter}, we present the method used to analyze the
QMAP experiment (the mapmaking process and the power spectrum extraction), 
as well as the combined results from both flights. 

\section{METHOD}

\subsection{From Scan Pattern to Map}

For each of the six channels,
the QMAP raw data set consists of $M$=35,318,400 observed data points 
\yi~ which we store in an $M$-dimensional vector $\bfy$. We subdivide
the mapped region into $N$ pixels whose centers $\bfri$ form a 
rectangular grid and let $x_i$ denote the true sky temperature in the
direction $\bfri$. Grouping these pixel temperatures into an 
$N$-dimensional map vector $\bfx$, we can write
\beq{a1}
    \bfy = \bfA \bfx + \bfn,
\eeq
where $\bfn$ denotes the random instrumental noise vector of size 
$M$ and $\bfA$ is a matrix of size $M \times N$ that encodes the QMAP 
scan strategy. We model $\bfn$ as a random variable with zero mean and 
covariance matrix given by $\bfN \equiv \langle \bfn \bfnt \rangle$. 
The scan strategy matrix $\bfA$ has 
$\bfA_{ij}=1$ if the $i^{th}$ observation points to the 
$j^{th}$ pixel, $0$ otherwise.
% if $N_i=j$ and 
% $\bfA_{ij}$=0 otherwise, where $N_i$ is the number of
% the pixel pointed to during the $i^{th}$ observation. 

The goal is to compute a map $\bfxT$ that estimates the true map $\bfx$ 
from the raw data $\bfy$. We use a linear method 
\beq{a2}
    \bfxT = \bfW \bfy,
\eeq
specified by some $N\times M$ matrix $\bfW$. Substituting (\ref{a1}) into 
(\ref{a2}) shows that the error in the recovered map is 
\beq{a3}
    \eta \equiv \bfxT - \bfx = [ \bfW \bfA - \bfI ] \bfx + \bfW \bfn,
\eeq
where $\bfI$ is the identity matrix. Choosing $\bfW$ to be (Tegmark 1997a, 
hereafter T97a)
\beq{a4}
    \bfW = [ \bfAt \bfM^{-1} \bfA ]^{-1} \bfAt \bfM^{-1}
\eeq
for some $M\times M$ matrix $\bfM$ gives $\bfW\bfA=\bfI$, 
so that $\bfxT=\bfx+\bfn$ can be interpreted as an honest-to-goodness map
where the pixel noise $\eta=\bfW\bfn$ is independent of $\bfx$.
The noise covariance matrix of the map is simply
\beq{a5}
    \Sig \equiv \langle \eta \etat \rangle = \bfW \bfN \bfWt.
\eeq

\subsection{Pre-whitening \& Pink Noise Removal}

Ideally, we would like to use the minimum-variance mapmaking method
% \footnote{
% A mapmaking method that retains all the cosmological information from the raw data.}
(Wright 1996; T97a), which corresponds to the choice $\bfM=\bfN^{-1}$.
In our case, the huge matrix $\bfN$ is far from diagonal, since long-term $1/f$ drifts 
(so-called pink noise) introduce strong correlations between the noise $n_i$ at 
different times. In other words, although direct application of equations (\ref{a2}) 
and (\ref{a5}) with $\bfM=\bfN^{-1}$ would give what we need in principle, inverting the 
non-sparse matrix $\bfN$ would require a Hubble time in practice.

Fortunately, we can obtain virtually the same answer by employing a series of numerical 
methods detailed in Tegmark (1997b, hereafter T97b). Since the statistical properties of the
noise are virtually constant in time, \ie, $\bfN$ is almost a 
circulant matrix\footnote{
A circulant matrix is one where each row is the previous 
one shifted one notch to the right. They are 
easily manipulated (inverted, diagonalized, \etc) with Fourier methods 
(see T97b).}, we replace the raw data $\bfy$ by a high-pass filtered data set
\beq{a6}
	\bfyT \equiv \bfD \bfy,
\eeq
where $\bfD$ is another circulant matrix 
(a convolution filter), chosen so that both 
$\bfD$ and the filtered noise covariance matrix $\bfNT\equiv\expec{\bfnT\bfnTt} = \bfD\bfN\bfDt$
are band-diagonal. 
Wright (1996) referred to this as ``pre-whitening'' 
and chose the filter so that $\bfNT\approx\bfI$.
In our case, however, $\bfNT$ is not quite circulant 
because of the omission of 
$\sim 600$ segments of calibration data (H98), 
but of the form $\bfNT=\bfNT_c+\bfNT_s$, 
where $\bfNT_c$ is circulant and the correction $\bfNT_s$ is extremely sparse.
We therefore choose $\bfM=\bfN_c^{-1}$, with $\bfyT$ as the new data set.

Defining $\bfAT \equiv \bfD \bfA$, we can rewrite (\ref{a1})
as $\bfyT$=$\bfAT \bfx + \bfnT$, so equations (\ref{a2}) and (\ref{a5}) now give
\beq{a11}
    \bfxT = [\bfA^t\bfD^t\bfM\bfD\bfA]^{-1} \bfA^t\bfD^t\bfM\bfD\bfy
\eeq
\beq{a12}
    \Sig  =  [\bfA^t\bfD^t\bfM\bfD\bfA]^{-1}  
             [\bfA^t\bfD^t\bfM\bfNT\bfM^t\bfD\bfA]
             [\bfA^t\bfD^t\bfM\bfD\bfA]^{-1}
\eeq
which can be evaluated on a workstation in about 24 hours.

\subsection{Offset Removal}

In order to make maps from the data, we need to remove instrumental 
offsets that are synchronous with the chopper position. Although
we find no evidence of atmospheric emission or sidelobe contamination,
there is thermal emission from the
instrument at the mK-level (H98).
%After removing calibration segments 
%from the data (H98), 
We solve for this scan-synchronous component by adding  
160 ``virtual pixels" $\bfx_2$ (corresponding to the 160 
sampling positions along the scan) to the map vector $\bfx_1$ and 
widening the scan strategy matrix $\bfA$ with an additional ``1'' in each row
in one of 160 extra columns. 
This allows us to rewrite equation (\ref{a1}) as
\beq{a13}
    \bfy = \bfA_{\rm 1} \bfx_{\rm 1} + \bfA_{\rm 2} \bfx_{\rm 2} = 
    \bfA \bfx + \bfn,
\eeq
where $\bfA_1$ and $\bfA_2$ are matrices of sizes $M \times N$ 
and $M \times 160$, respectively, so $\bfx$ becomes a vector of dimension
$N+160$ and $\bfA$ a matrix of size $M \times (N+160)$.  
Since the scan strategy is so well interconnected, this method 
is able to produce an offset-free map at the cost of only a marginal increase
in pixel noise. However, we do not wish to assume that the offset remains
constant during the entire flight, as even a 10\% variation in a 
1 mK offset would cause an artificial signal comparable to the CMB.
Since our offset is almost entirely localized in frequency to the
first two harmonics of the scan rate (H98), we therefore combine the
virtual pixel method with an extremely conservative approach illustrated in 
Figure 1: we make $\bfD$ a notch filter which annihilates all signals at these 
two frequencies, as well as the DC (0 Hz) component.
Our filtering is thus more of a ``pre-blueing'' than a pre-whitening.
The price we pay for this conservatism is that we lose essentially all 
information about the CMB dipole, which was detected.

\bigskip
\centerline{{\vbox{\epsfxsize=9.5cm\epsfbox{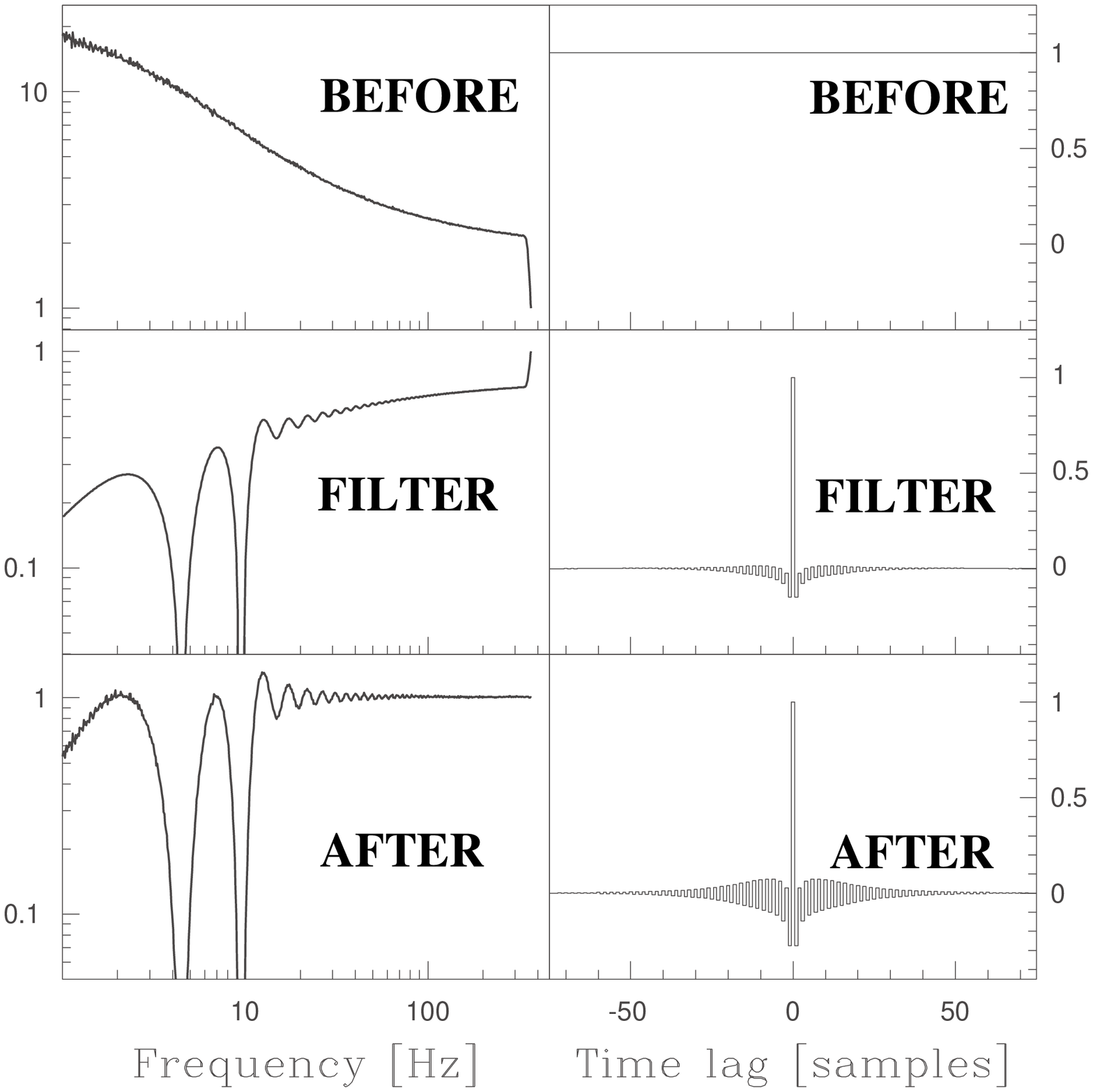}}}}
\figcaption{The filtering procedure is illustrated in
the frequency domain (left) and in the time domain
(right) for the Ka1 channel of flight 2.
The noise power spectrum (top left) contains a $1/f$-component
which causes the noise to be almost perfectly correlated between
measurements close together in time (top right).
By multiplying the Fourier transformed signal
by an appropriate filter
(middle left), which corresponds to applying a
convolution filter to the signal (middle right),
we obtain 
filtered data $\bfyT$ that has a white noise power
spectrum with three ``notches'' (lower left), corresponding to a
block-diagonal time autocorrelation function 
% $\expec{\tilde{n}_i\tilde{n}_j}\approx\delta_{ij}$ 
(lower right).
1 sample $\approx 1.36$~ms. Vertical units are arbitrary.
}
%\bigskip

\subsection{Combining maps}

When combining two maps $\bfx_1$ and $\bfx_2$ 
of the same angular resolution
into a single map $\bfxT$, we use the minimum-variance weighting 
\beq{ComboEq1}
\bfxT = \left[\Sig_1^{-1}+\Sig_2^{-1}\right]^{-1}
\left[\Sig_1^{-1}\bfx_1 + \Sig_2^{-1}\bfx_2\right].
\eeq
The resulting covariance matrix for
$\bfxT$ is therefore
\beq{CombiEq2}
\Sig =\left[\Sig_1^{-1}+\Sig_2^{-1}\right]^{-1}.
\eeq
When combining maps of different resolution, the one with the
higher resolution was first smoothed to the lower resolution.

% When combining two maps $\bfx^{(1)}$ and $\bfx^{(2)}$ 
% of the same angular resolution
% into a single map $\bfxT$, we used a separate minimum-variance weighting 
% for each pixel, \ie,
% \beq{LuigiEq1}
% x_i = w^{(1)}_i x^{(1)}_i + w^{(2)}_i x^{(2)}_i
% \eeq
% with weights 
% $w^{(k)}_i\propto 1/\Sig^{(k)}_{ii}$ normalized so that
% $w^{(1)}_i+w^{(2)}_i=1$. The resulting covariance matrix for
% $\bfxT$ is therefore 
% \beq{LuigiEq2}
% \Sig_{ij}=[w^{(1)}_i]^2\Sig^{(1)}_{ij} 
%           +[w^{(2)}_i]^2\Sig^{(2)}_{ij}.
% \eeq
% When combining maps of different resolution, the one with the
% higher resolution was first smoothed to the lower resolution.

\subsection{Wiener-Filtered Map}

In addition to the $\bfxT$ map, we also compute the $Wiener$ map
$\bfx_{\it w}$ given by (T97a)
\beq{a14}
    \bfx_{\it w} = \bfS [\bfS + \bfN]^{-1} \bfxT,
\eeq
where $\bfS$=$\langle \bfx \bfxt \rangle$ is the CMB covariance matrix,
defined as
\beq{a15}
    \bfS_{ij} = \sum_{\ell=2}^{\infty}
                           \frac{(2 \ell + 1)}{4 \pi}
	                   P_{\ell} (\bfrh_i\cdot\bfrh_j) 
	                   B_{\ell}^2 C_{\ell}.
\eeq
To avoid imprinting features on any particular scale, we use a 
flat fiducial power spectrum $C_{\ell}$ normalized to 
$Q=30\microK$, which is roughly the CMB power level we find in the maps. 
We approximate the QMAP beam by a circular 
Gaussian with FWHM values given by D98, \ie, 
     $B_{\ell} \approx e^{-\theta^2 \ell(\ell+1)/2}$ and 
     $\theta \equiv \sqrt{8 \ln 2}$ FWHM. 
The Wiener filtered map $\bfx_{\it w}$ contains the same information as 
$\bfxT$, but it is more useful for visual inspection since it is less noisy.

\subsection{From Map to Power Spectrum}

The signal-to-noise (\sn) method (Bond 1995; Bunn and Sugiyama 1995) compresses 
the information content of a CMB map into a vector $\bfz \equiv \bfBt \bfxT$,
where $\bfB$ is an $N\times N$ matrix whose $i^{th}$ column satisfies the generalized 
eigenvalue equation
\beq{a17}
    \bfS \bfb_i = \lambda_i \Sig \bfb_i,
\eeq
normalized so that $\bfbt_i\Sig\bfb_i =1$ and sorted by decreasing $\lambda_i$.
As described in {\eg} T97b, the $N$ numbers $z_i$ are uncorrelated, \ie, 
\beq{a18}
    \langle z_i z_j \rangle =
    \bfbt_i (\Sig + \bfS) \bfb_j =
    [{\rm 1}+\lambda_i] \delta_{ij},
\eeq
and their variance $\langle z_i^2\rangle$ has a contribution of $1$ from noise and 
$\lambda_i$ from signal. This means that the eigenvalue $\lambda_i$ can be 
interpreted as a {\sn} ratio for $z_i$, and the quantities 
$q_i\propto (z_i^2-1)$ can be used as estimators of the power spectrum
$\delta T_\l\equiv\ell (\ell+1) C_{\ell}/ 2 \pi$, since 
$\expec{q_i}\propto\sum_\l W_\l\delta T_\l$ for some window function 
$W_\l$. The $q_i$ tend to probe smaller scales as $i$ increases and 
the {\sn} drops. The band power measurements in Table 1 and Figure 3 have
been computed by normalizing the individual $q_i$ so that their window functions
integrate to unity and then averaging them in bands with a
minimum-variance weighting, to minimize error bars. 
  
\section{DATA ANALYSIS}

\subsection{Pipeline Tests}

We tested our data analysis pipeline by generating mock raw data sets
that incorporate the QMAP scan strategy. These mock data sets were
processed though the pipeline, recovering the original maps. 
When adding Monte Carlo white and pink noise to these mock data sets, we recovered 
maps with pixel noise consistent with the noise covariance matrix $\Sig$ 
computed by the pipeline. Likewise, when adding scan-synchronous offsets 
to these mock data sets, we recovered the original maps as well as the 
offsets. As expected, the original maps were faithfully recovered even when the 
first harmonics of these mock offsets were varied slowly 
throughout the flight.

We repeated our analysis for a range of pixel sizes. As expected, 
we found that as long as the pixels were smaller than the Shannon oversampling 
limit (about 2.5 times smaller than the FWHM), the maps $\bfxT$ were virtually 
independent of the pixelization.

We made $\bfNT$ as band-diagonal as possible using a filter
$\bfD$ of band width $L$, then neglected the tiny elements of 
$\bfNT$ further than $L/2$ from the diagonal.
Tests with increasing $L$-values showed that the results converged
for $L\sim 150$, so we used $L=320$ in the analysis to be conservative.

To test whether the Wiener maps were sensitive to our choice of 
power spectrum normalization, we generated Wiener maps for fiducial 
power spectra with $Q$=20, 30 and 40$\microK$. The visual difference 
between the two extreme normalizations was minimal: the maps had
the same spatial features in the same locations, the 20$\microK$ map simply
being slightly smoothed relative to the 40$\microK$ map.

Finally,
if the beam size $\theta$ is overestimated by 1\%, the 
band power $\delta T_\ell$ is overestimated by
$[(\theta\ell)^2-2]$ percent. The first term comes from the above-mentioned
Gaussian beam correction $B_\ell$ and the second from the calibration, which 
involves the beam area $\propto\theta^2$.
Repeating our full analysis with the assumed FWHM reduced by 
$1\sigma$ ($\sim 3\%$), the first effect 
decreases the normalization of the two
combined Ka band powers in Table 1 by 0.3\% and 4\%, respectively,
whereas the second effect of course gives a 6\% increase.
% combined Ka band powers in Table 1 by 0.5\%, and 6\%, respectively.

\subsection{Data Tests}

The above-mentioned scan-synchronous offset was around 1mK (FL2) 
and 10mK (FL1) peak-to-peak. Although our notch filter technique immunized 
the results towards drifts in this offset, no such drifts were 
actually detected.

How statistically significant is our detection of signal in the maps?
Consider the null hypothesis that a map $\bfxT$ contains merely noise,
\ie, $\expec{\bfxT \bfxTt}=\Sig$. Given the alternative hypothesis 
$\expec{\bfxT \bfxTt}=\Sig+\bfS$, one can show that the most powerful
``null-buster'' test for ruling out the null hypothesis is using the generalized 
$\chi^2$-statistic
\beq{NullbusterEq}
    \chi^2 \equiv {\bfxTt\Sig^{-1}\bfS\Sig^{-1}\bfxT - \tr\Sig^{-1}\bfS
    \over
    \left[2\tr\left\{\Sig^{-1}\bfS\Sig^{-1}\bfS\right\}\right]^{1/2}},
\eeq
which can be interpreted as the number of ``sigmas'' at which the
null noise-only hypothesis is ruled out. The results of this test are given for 
all the individual maps in D98 and H98, and show that signal is 
detected at significance levels above $15\sigma$ in both flights.

Of these 11 maps (see D98 and H98), many have a substantial spatial overlap.
This allows a series of powerful consistency tests, since many potential
contaminants affect data from different bands and polarization channels differently.
As detailed in D98 and H98, we applied the same null-buster test to
the difference of the various maps in each band where they overlap spatially, 
and in all cases found the difference maps consistent with pure noise.
Our S/N eigenmode analysis gives the same conclusion: 
significant signal in the best eigenmodes of the individual 
channels, but S/N-coefficients consistent with pure noise in the 
difference maps.
Thus all of the significant signal appears to be common to 
the different channels, indicating that the bulk of the detected signal is
due to temperature fluctuations on the sky.

\subsection{Foreground Contamination}

To constrain the frequency dependence of our signal, we repeated
the null-buster test for weighted difference maps of the form
\beq{foreground}
   \bfxT \equiv \bfxT_1-(\nu_2/\nu_1)^{\beta}\bfxT_2, 
\eeq
where the map $\bfxT_1$ and the frequency $\nu_1$ was for the Ka-band 
and $\bfxT_2$ and $\nu_2$ was for the Q-band.
This placed a $2-\sigma$ lower limit on the spectral 
index $\beta$ of $-1.4$, 
which means that the signal cannot be dominated by foregrounds such as
free-free emission ($\beta \sim-2.15$) or synchrotron radiation 
($\beta \sim -2.8$). A more detailed foreground analysis,
cross-correlating the maps with various foreground templates, 
will be presented in a separate paper (de Oliveira-Costa {\etal} 1998).

%  \bigskip
%  \bigskip

   \vskip-2.3cm
   \hglue-0.1cm
   \centerline{\vbox{\epsfxsize=11cm\epsfbox{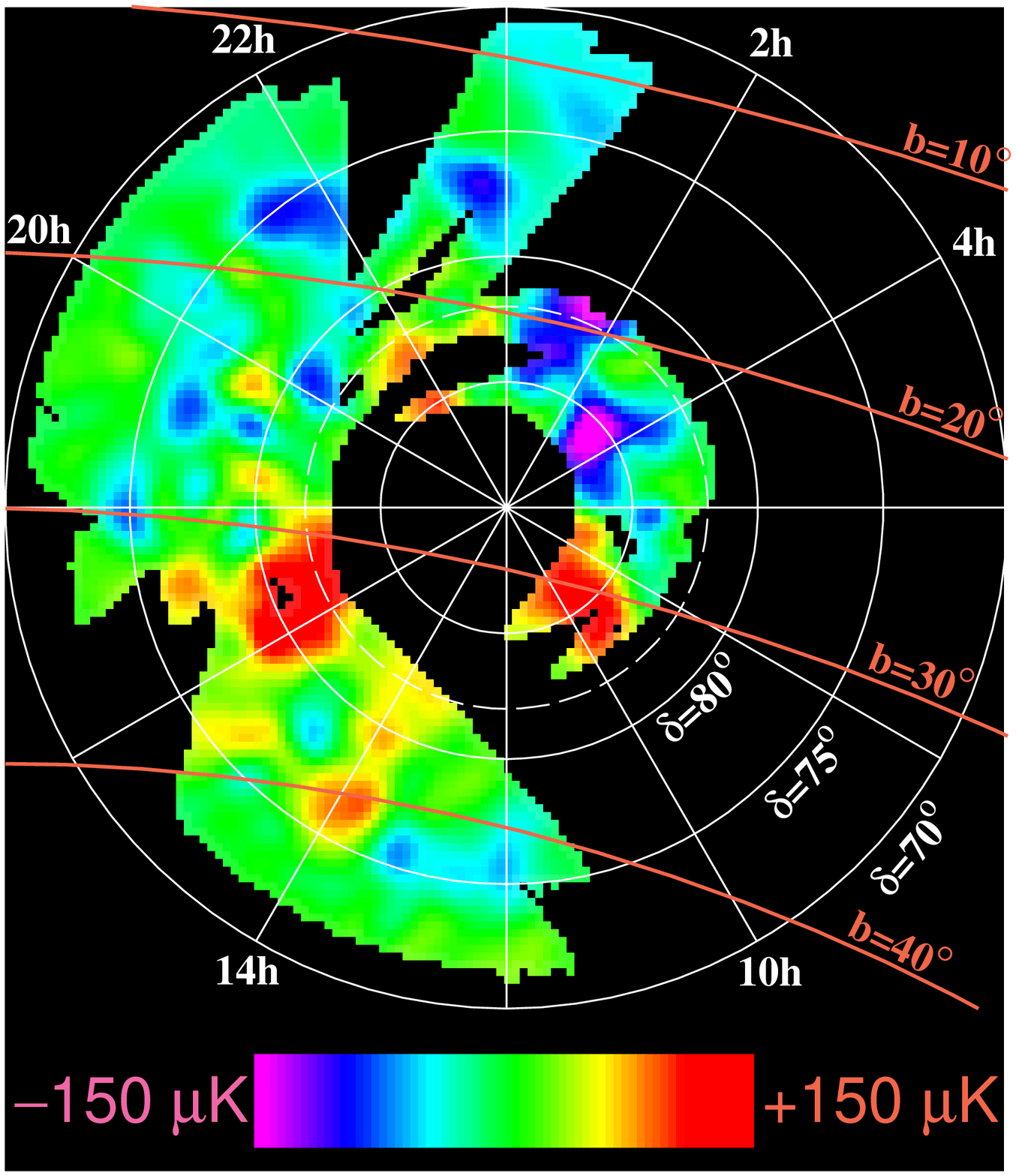}}}
   \vskip-2cm
   \figcaption{Wiener-filtered map of the QMAP combined data. 
   The dashed circle shows the Saskatoon coverage.
%    The CMB temperature is shown in coordinates where the NCP is near 
%    the center, with RA being zero at the top and increasing clockwise.
   \label{MapFig}
   }

\section{Results \& Conclusions}

The Wiener map obtained by combining all the data from both flights 
is shown in Figure~2. The above-mentioned generalized $\chi^2$-test
shows that the signal observed in this figure is significant at the 
$\simgt 15\sigma$-level.

This map covers 527 square degrees, and has a substantial overlap 
with the 200 square degree Saskatoon map. Visual comparison of the two maps in 
the overlap region reveals striking similarities, providing further indication
that the bulk of the detected signal is due to temperature fluctuations on the 
sky rather than systematic effects. A  detailed statistical comparison of the 
QMAP and Saskatoon data sets will be presented in a future paper.

   \bigskip
%   \vskip-1.5cm
%   \centerline{\vbox{\epsfxsize=9.0cm\epsfbox{angelica_powerfig.eps}}}
%   \vskip-1.5cm
   \vskip-0.2cm
   \centerline{\vbox{\epsfxsize=8.0cm\epsfbox{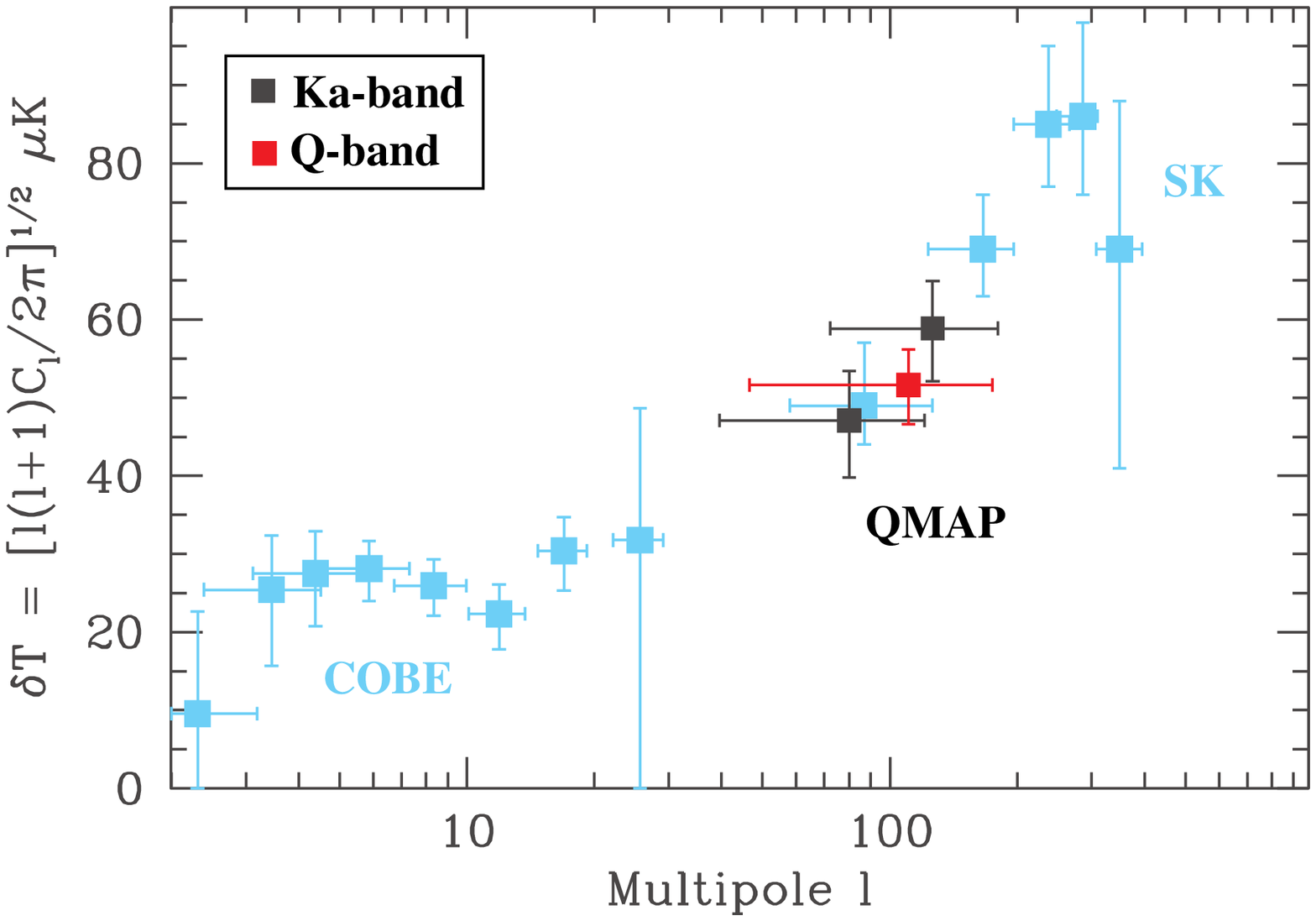}}}
   \vskip-0.3cm
   \figcaption{Angular power spectrum of combined data set
   shown in Figure~2. The results are listed in Table~1.}
   \bigskip

The power spectrum for the total data set (FL1+FL2) is given in 
Figure~3 and Table~1.
We see that it agrees well with the Saskatoon power spectrum (Netterfield {\etal} 1997),
showing a rise on degree scales to power levels substantially above those 
found on large scales by COBE. 
The calibrated raw data with pointing will be made public after
publication of this {\it Letter}.

   \bigskip
   % \tabcaption{The Angular Power Spectrum}
   {\footnotesize
   Table~1. --- The angular power spectrum.
   The band powers $\delta T_\l\equiv[\l(\l+1)C_\l/2\pi]^{1/2}$ 
   have window functions whose mean and 
   rms width are given by $\l_{eff}$ and $\Delta\l$.
   The full window functions are available at   
   {\it http://dept.physics.upenn.edu/cmb.html}
   and {\it http://pupgg.princeton.edu/$\sim$cmb/welcome.html}.
%  The errors $\delta T$ include both noise and sample variance, 
%  and are uncorrelated within each pair of Ka-points.
   The errors $\delta T$ are uncorrelated within each pair of 
   Ka-points. The first and and second one in each pair is
   dominated by sample variance and detector noise, respectively. 
   Calibration errors are not included (see H98).
   \begin{center}
   \begin{tabular}{lcrcl}
   \hline
   \hline
   \multicolumn{1}{l}{Flight}             	  &
   \multicolumn{1}{l}{Band}             	  &
   \multicolumn{1}{c}{$\l_{eff}$}          & 
   \multicolumn{1}{c}{$\Delta \l $}  & 
   \multicolumn{1}{c}{$\delta T$}    \\
   \hline
    1    &Ka	&92	&45  &$49^{+6}_{-7}$\\
         &Q	&84	&46  &$47^{+8}_{-10}$\\
   \hline
    2    &Ka	&91	&47  &$46^{+10}_{-12}$\\
         &Ka	&145	&64  &$63^{+10}_{-12}$\\
         &Q	&125	&67  &$56^{+5}_{-6}$\\
   \hline
    1+2  &Ka	&80	&41  &$47^{+6}_{-7}$\\
	 &Ka	&126	&54  &$59^{+6}_{-7}$\\
         &Q	&111	&64  &$52^{+5}_{-5}$\\
   \hline
   \end{tabular}
   \end{center}
   }
   \bigskip

We would like to thank Wayne Hu for helpful comments.
This work was supported by
a David \& Lucile Packard Foundation Fellowship (to LP),
a Cottrell Award from Research Corporation, an NSF NYI award, 
NSF grants PHY-9222952 and PHY-9600015, 
NASA grant NAG5-6034 and Hubble Fellowships 
HF-01044.01$-$93A (to TH) and HF-01084.01$-$96A (to MT)
from by STScI, operated by AURA, Inc. under NASA contract NAS5-26555.

%%%%%%%%%%%%%%%%%%%%%% REFERENCES: %%%%%%%%%%%%%%%%%%%%%%%%%

% \clearpage
% \setcounter{secnumdepth}{0}
% \section{REFERENCES}

%%%%%%%%%%%%%%%%%%%%%% FIGURES: %%%%%%%%%%%%%%%%%%%%%%%%%%%%%
%%%%%%%%%%%%%%%%%%%%%% TABLES: %%%%%%%%%%%%%%%%%%%%%%%%%%%%%%

\end{document}